\documentclass[manuscript]{acmart}

\AtBeginDocument{%
  \providecommand\BibTeX{{%
    \normalfont B\kern-0.5em{\scshape i\kern-0.25em b}\kern-0.8em\TeX}}}

\copyrightyear{2023} 
\acmYear{2023} 
\setcopyright{rightsretained} 
\acmConference[MuC '23]{Mensch und Computer 2023}{September 3--6, 2023}{Rapperswil, Switzerland}
\acmBooktitle{Mensch und Computer 2023 (MuC '23), September 3--6, 2023, Rapperswil, Switzerland}\acmDOI{10.1145/3603555.3608551}
\acmISBN{979-8-4007-0771-1/23/09}

\usepackage{verbatim}

\usepackage{enumitem}

\usepackage{multirow}

\usepackage{graphicx}
\usepackage{textcomp}
\graphicspath{ {./images/} }

\usepackage[utf8]{inputenc}
    
\usepackage{tabularx}
\usepackage{booktabs}

\newcolumntype{Y}{>{\centering\arraybackslash}X}
\newcolumntype{R}{>{\raggedleft\arraybackslash}X}
\newcolumntype{L}{>{\raggedright\arraybackslash}X}
    
\usepackage{hyperref}

\usepackage{appendix}

\usepackage{xspace}
\newcommand*{\eg}{e.g.,\@\xspace}
\newcommand*{\ie}{i.e.,\@\xspace}

\begin{document}

\title[Identifying Explanation Needs of End-users]{Identifying Explanation Needs of End-users: Applying and Extending the XAI Question Bank}


\author{Lars Sipos}
\authornote{Both authors contributed equally to this research.}
\email{lars.sipos@fu-berlin.de}
\orcid{0009-0000-0572-6490}
\author{Ulrike Sch\"afer}
\authornotemark[1]
\email{ulrike.schaefer@fu-berlin.de}
\orcid{0000-0002-9070-1665}
\affiliation{%
  \institution{Freie Universit\"at Berlin}
  \streetaddress{Königin-Luise-Str. 24/26}
  \city{Berlin}
  \country{Germany}
  \postcode{14195}
}

\author{Katrin Glinka}
\email{katrin.glinka@fu-berlin.de}
\orcid{0000-0002-4232-8907}
\affiliation{%
  \institution{Freie Universität Berlin}
  \streetaddress{Königin-Luise-Str. 24/26}
  \city{Berlin}
  \country{Germany}
  \postcode{14195}
}

\author{Claudia M\"uller-Birn}
\orcid{0000-0002-5143-1770}
\affiliation{%
  \institution{Freie Universit\"at Berlin}
  \streetaddress{Königin-Luise-Str. 24/26}
  \city{Berlin}
  \country{Germany}
  \postcode{14195}
}
\email{clmb@inf.fu-berlin.de}

\renewcommand{\shortauthors}{Sipos and Sch\"afer et al.}

\begin{abstract}
Explainable Artificial Intelligence (XAI) is concerned with making the decisions of AI systems interpretable to humans. Explanations are typically developed by AI experts and focus on algorithmic transparency and the inner workings of AI systems. Research has shown that such explanations do not meet the needs of users who do not have AI expertise. As a result, explanations are often ineffective in making system decisions interpretable and understandable. 
We aim to strengthen a socio-technical view of AI by following a Human-Centered Explainable Artificial Intelligence (HC-XAI) approach, which investigates the explanation needs of end-users (\ie subject matter experts and lay users) in specific usage contexts. 
One of the most influential works in this area is the XAI Question Bank (XAIQB) by Liao et al. The authors propose a set of questions that end-users might ask when using an AI system, which in turn is intended to help developers and designers identify and address explanation needs. Although the XAIQB is widely referenced, there are few reports of its use in practice. 
In particular, it is unclear to what extent the XAIQB sufficiently captures the explanation needs of end-users and what potential problems exist in the practical application of the XAIQB. 
To explore these open questions, we used the XAIQB as the basis for analyzing 12 think-aloud software explorations with subject matter experts, \ie art historians. We investigated the suitability of the XAIQB as a tool for identifying explanation needs in a specific usage context. 
Our analysis revealed a number of explanation needs that were missing from the question bank, but that emerged repeatedly as our study participants interacted with an AI system. We also found that some of the XAIQB questions were difficult to distinguish and required interpretation during use. 
Our contribution is an extension of the XAIQB with 11 new questions. In addition, we have expanded the descriptions of all new and existing questions to facilitate their use. We hope that this extension will enable HCI researchers and practitioners to use the XAIQB in practice and may provide a basis for future studies on the identification of explanation needs in different contexts.
\end{abstract}



\maketitle

\section{Introduction}
In recent years, due to increasing model complexity and the use of non-transparent black-box AI models, the field of Explainable Artificial Intelligence (XAI) has emerged to make AI decisions understandable and interpretable~\cite{saeedExplainableAIXAI2023, ehsan2020}.
%
However, AI explanations are often developed by and for AI experts (\eg~\cite{Cai2019, Hohman2019, Madaio2020}) and require advanced AI knowledge that many users lack~\cite{ehsan2021}. Studies that consider users with little or no AI expertise often rely on crowdsourcing platforms to analyze the suitability of explanations (\eg~\cite{Ribeiro2016, Cheng2019, Yu2020}), which decontextualizes interpretability~\cite{Benjamin_ExplStrategies2022}.
As a result, research has emphasized that explanations often do not meet the needs of users without AI expertise, instead, they serve, for example, AI engineers and data scientists~\cite{ehsan2021, Liao2020}.
%
In light of this situation, the field of Human-Centered Explainable Artificial Intelligence (HC-XAI) seeks to make model decisions understandable by studying the explanation needs, \ie XAI needs, of users who lack expertise in AI (see, \eg~\cite{ehsan2020, kim2023, Liao2020,liao2022humancentered, EhsanUpol2022}).
%
Different approaches were proposed to capture and characterize XAI needs. However, some approaches see users just as one part of many stakeholders~\cite{preece2018stakeholders, langer2021, suresh2021a}. In addition, even when the user is the focus of the examination, a theory-driven conceptual framework has often been developed without direct empirical evaluation~\cite{Mohseni2021, preece2018stakeholders, langer2021, suresh2021a, sanneman2020, Wang2019}. Rarely have interviews been used to subsequently explore user needs with the developed framework~\cite{Hong2020, Liao2020}.

%
One of the most influential works in this area is the XAI Question Bank (XAIQB) by Liao et al.~\cite{Liao2020}, which provides a human-centered approach to eliciting explanation needs that can be adapted to different usage contexts. The XAIQB consists of a set of questions that summarize user needs in terms of explainability. It is one of the most referenced works on XAI, indicating the importance of the topic and the relevance of its contribution\footnote{434 citations according to Google Scholar (May, 2023) and 214 according to ACM digital library}.  
%
However, it is unclear whether and to what extent the XAIQB questions sufficiently capture the actual explanation needs of end-users~\cite{sovrano2022a}, \ie subject matter experts and lay users~\cite{saeedExplainableAIXAI2023}. 
Few studies directly use or apply the XAIQB or reflect on its applicability, and even fewer empirically test the validity of the questions in user studies (see~\autoref{sec:theory}). 

We aimed to further explore and evaluate the applicability of the XAIQB, specifically in terms of whether and how it can be used as a tool to uncover the explanation needs of end-users.
%
We, therefore, conducted a qualitative study using the XAIQB as a tool to systematically capture the explanation needs of end-users. As part of our ongoing research on the potential for human-AI collaboration, we conducted a software exploration of a computer vision-based image retrieval system with 12 art historians using a think-aloud approach~\cite{Lewis_1982}. We observed and recorded our participants as they interacted with the AI system, and then used the XAIQB to qualitatively identify our participants' explanation needs.
This resulted in four main contributions:
(1) we explore and describe an approach for applying the XAIQB with end-users,
(2) we propose additional end-user-centered questions to extend the XAIQB, 
(3) we provide refined definitions and examples for all XAIQB questions, and
(4) we discuss the scope and applicability of the XAIQB.

\section{Explainable AI Question Bank} \label{sec:theory}
Liao et al.~\cite{Liao2020} propose the XAIQB to explore user needs for understanding AI in terms of questions a user might ask about the AI. 
The authors were guided by conceptual work such as Lim and Deys'~\cite{Lim2010} taxonomy to support intelligibility in context-aware applications. Liao et al. extended Lim and Deys'~\cite{Lim2010} categories by mapping them to explanation methods and algorithms. 
This resulted in nine explainability needs categories of prototypical questions that users may have about an AI system: \textit{Input, Output, Performance, How (global), Why, Why not, What if, How to be that, How to still be this} and \textit{Others}~\cite{Liao2020}. Based on an initial set of questions, Liao et al. conducted 20 interviews with UX and design practitioners at IBM working on AI products to explore gaps in XAI algorithm work and practices. During the interviews, they did not consider the direct application of an AI system but instead focused on the practitioners' prior experience with user needs and feedback on the question bank. However, as Liao et al. acknowledge, users ``may not have a deep technical understanding of AI, but hold preconceptions of what constitutes useful explanations for decisions made in a familiar domain''~\cite{Liao2020}. Thus, in our research, we wanted to focus solely on the explanation needs of end-users. 

As a first step, we wanted to understand how other researchers or practitioners were using the XAIQB.
Many papers we found in our literature search only mention Liao et al.~\cite{Liao2020} as part of a literature review (\eg \cite{antoniadi2021, chromik2021, luna2020, kaluarachchi2021}) or for theoretical arguments (\eg \cite{drozdal2020, poursabzi-sangdeh2021, shneiderman2020}). 
A few studies used the XAIQB as a theoretical basis.
Nguyen et al.~\cite{nguyen2022a} developed a conversational agent to inform users about explanations in a manner similar to human-to-human conversations. The authors extended the XAIQB with quality-controlled paraphrases and explored appropriate XAI methods to answer these questions~\cite{nguyen2022a}. Similarly, Shen et al.~\cite{shen2021a} summarized common forms of natural language processing explanations and compared them to the XAIQB~\cite{shen2021a}. Liao et al.~\cite{liao2021question} developed a question-driven design process, including a mapping guide between user questions and XAI requirements based on the XAIQB, 21 interviews with AI practitioners, and two healthcare use cases~\cite{liao2021question}.
We could identify only a few papers that use the XAIQB to empirically investigate user needs. Kim et al.~\cite{kim2023} used the XAIQB to develop open-ended questions and a survey to explore how explainability can improve human-AI interaction. Twenty users were surveyed about their needs for a real-world AI bird identification application. In addition to the nine categories of Liao et al.~\cite{Liao2020}, the question category \emph{transparency} was added on a theoretical basis~\cite{kim2023}. He et al.~\cite{he2023} developed a user needs library based on the XAIQB, which they adapted to a medical context based on their literature review. They then designed an XAI prototype for the medical domain and analyzed the needs and preferences of consumer users with respect to explanations~\cite{he2023}.
However, to the best of our knowledge, there is no study that empirically evaluated the XAIQB and its applicability for end-users directly.

\section{Methods}

The data was collected as part of an ongoing study on human-AI collaboration in humanities research\footnote{All information about the complete study, including the semi-structured interviews, can be found in Glinka et al.~\cite{glinka2023criticalreflective}. The analysis and our contributions presented in this submission to MuC have a different focus and have not been submitted or published elsewhere.}. For this study, we recruited a total of 12 art historians to participate in our software exploration study in March and April 2022. 
Quota sampling was used based on participants' professional backgrounds and qualitatively assessed familiarity with computational approaches.
Eligibility criteria included having at least one academic degree in art history and working in an art history-related profession.
We conducted a semi-structured interview followed by the software exploration to gain direct insight into our participants' usage practices and explanation needs. Only the software exploration part was used for the analysis presented in this paper. 
During the think-aloud software exploration, participants used the computer vision-based image retrieval system \texttt{imgs.ai}~\cite{imgsai2023}\footnote{More information is available at \url{https://imgs.ai/}.} to perform a real-world image retrieval task based on their current research. 
The tool allows users to perform a visual search on a set of museum collection records and allows them to interactively select different embeddings, \ie compressed semantic descriptors for images, and a distance metric as a similarity criterion when performing their search~\cite{imgsai2023}. The tool does not require advanced technical knowledge\footnote{We are not involved in the development of \texttt{imgs.ai}.}. 

The software explorations lasted approximately 25 minutes each, were conducted in German, and were recorded.
During the interaction, we did not encourage study participants to elaborate on their need for explanation. Instead, the software exploration mimicked a real-world image retrieval task in which participants expressed their intrinsically motivated needs for explanation in the context of use.
We transcribed the audio recordings of the software exploration and consulted the video recordings of the users' screens to additionally transcribe relevant actions that were not verbalized. 

Our analysis was conducted by three of the authors and consisted of three distinct phases: \emph{preparation}, \emph{coding}, and \emph{refinement}. In the preparation phase, we reviewed and discussed all of the XAIQB questions. For each question, we considered how it might be interpreted in general and specifically in our context. For example, we determined that the question ``What is the source of the data?'' applies only to training data and not to images that can be selected in the tool. Based on our discussions, a working definition was developed for each question.

We used MAXQDA\footnote{\url{https://www.maxqda.com/}} to code the transcribed software explorations. We applied a dual coding approach, \ie we defined all 49 XAIQB questions as codes that we would assign deductively. In addition, we inductively assigned additional codes that we initially grouped under ``new question''. We performed the coding in an iterative process consisting of four rounds. In each round, three of the authors independently coded the same set of transcribed software explorations (two explorations in round 1, three in rounds 2 and 3, and four in round 4). After each round, we met to discuss the assigned codes and to explain why we applied a code to a particular segment. Likewise, we specified ``new questions'' and discussed why that question should be added to the XAIQB. During our discussions, we went through each code assignment and debated until a consensus was reached. We used our notes from the discussion to iteratively refine the question definitions from the preparation phase. For each of the ``new question'' segments, we elaborated on why this question expressed an explanation need that was different from those in the XAIQB. If a consensus was reached, the wording of the new question was refined, assigned a category, given a definition, and added as a code for the next rounds.
Finally, in the refinement phase, we performed a plausibility check on each newly added question to evaluate whether it was relevant to other AI systems, not just the specific tool we used for the software exploration. We removed questions that did not pass this check. We also reviewed all the question definitions again, refining them for clarity and adding application examples.

\section{Results}
Our analysis led to the identification of 11 new questions that represent the explanation needs expressed by our participants during the operation of the AI system, and a new ``need category'' for the XAIQB, see \autoref{tab:new-questions}. 
As with our new questions, we have developed refined definitions and examples to clarify and improve the applicability to all 49 other XAIQB questions as part of the analysis process.
The full table can be found in the supplementary material or on OSF\footnote{\url{https://osf.io/vu9zg}}.
During software exploration, it became clear that the AI system was not self-explanatory, and users might have questions about its user interface. Thus, we introduced the new category of \emph{UI Questions}. For example, users were unsure about the system's affordances, which are addressed by the newly introduced \emph{Question 1} ($Q1$), and the functionality of specific UI elements ($Q2$). We suggest adding the existing XAIQB question ``Other: What does [ML terminology] mean?'' into this new \emph{UI} category, since users interacting with an AI-based tool would only be exposed to ML terminology if it is integrated into the interface. 
We added two questions $Q3$ and $Q4$ to the existing categories \emph{How} and \emph{What if}. The new questions focus on system parameters, \ie values that users can interact with within the system, and how they affect the system or change the prediction, as opposed to the initial XAIQB which only addresses the hyperparameters set by the developers.
Questions $Q5$ through $Q8$ extend the category \emph{Output}, specifically in the context of AI-based information retrieval or recommender systems. Our participants repeatedly showed interest in the source, scope, and amount of output data. The XAIQB already included the question ``Output: What kind of output does the system give?'', which we interpreted as ``asking about the type of output, e.g. images, text, probabilities, classifications.''. Therefore, question $Q6$ is needed to specify the scope, since ``kind'' does not include characteristics of possible output data, \eg text from a limited range of genres. Several participants asked about the number of images from a particular museum that could be recommended. Therefore, question $Q7$ addresses the amount of output data and is equivalent to the existing XAIQB question ``Input: What is the sample size?'', which focuses only on the amount of data used for training. In addition, participants expressed interest in the logic behind the initial output ($Q8$), which are the items displayed before the user intentionally interacts with a system, such as movies recommended in a streaming service before users specify their preferences. This need arose because participants felt confused when the system did something without their input.
Finally, we added three new questions to the \emph{Others} category. We have included questions $Q9$ and $Q10$ in the subcategory \emph{Others (change)}, as they are a more specific version of the already included question ``How/why will the system change/adapt to/improve/drift over time? (change)''. Participants expressed the need to know if their own interaction will be used for further development of the system ($Q10$). This need relates to questions such as ``Do the developers benefit from my interaction? Will my data be used or integrated into the system?''. Question $Q9$ asks directly about how updates have already affected the system, rather than asking about future changes.
Our last proposed question ($Q11$) asks about the authors, i.e. the developers and designers of the tool. This question was asked frequently during our interviews and relates to the perceived reliability and responsibility of the tool's authors.

\begin{table*}[t]
\resizebox{\textwidth}{!}{\begin{tabular}{@{}ll|l|l@{}}
\toprule
No. &
  Category &
  Question &
  Definition \\ \midrule
1 &
  UI &
  What are the affordances of the system? &
  \begin{tabular}[c]{@{}l@{}}Asking about possible actions of the system, e.g., ``Can this system work\\  with my files? Am I able to upload my own pictures for prediction?''.\end{tabular} \\ \midrule
2 &
  UI &
  What does this UI element do? &
  \begin{tabular}[c]{@{}l@{}}Asking about the functionality of a specific UI element, e.g., ``What \\ happens when I click this button?'' or ``Does this clear my selection?''.\end{tabular} \\ \midrule
3 &
  How &
  How does {[}parameter X{]} affect the system? &
  \begin{tabular}[c]{@{}l@{}}Asking how a system parameter, i.e., a value changeable by the user, is \\ affecting the prediction behavior. For example, ``How does embedding X \\ influence the prediction'' or ``What does a learning rate of X do?''.\end{tabular} \\ \midrule
4 &
  What-if &
  What would the system predict if the parameter changes from P to Q? &
  \begin{tabular}[c]{@{}l@{}}Asking about a change in the outcome if a specific system parameter's \\ value is updated e.g., ``What is the prediction when I change the \\ learning rate from P to Q?''\end{tabular} \\ \midrule
5 &
  Output &
  What is the source of the information object(s)? &
  \begin{tabular}[c]{@{}l@{}}Asking about the source of the recommended information object(s), i.e., a \\ well-defined identifiable instance of an output. For example, ``From \\ which museum is the recommended painting?''.\end{tabular} \\ \midrule
6 &
  Output &
  What is the scope of the output data? &
  \begin{tabular}[c]{@{}l@{}}Asking about the scope of the output data, \\ e.g., ``Are the recommendations only from specific time frames?''.\end{tabular} \\ \midrule
7 &
  Output &
  What is the amount of the output data? &
  \begin{tabular}[c]{@{}l@{}}Asking about the amount of the output data, \\ e.g., ``How many images can this system recommend?''.\end{tabular} \\ \midrule
8 &
  Output &
  What logic is used for the initial output/recommendation? &
  \begin{tabular}[c]{@{}l@{}}Asking about why a specific recommendation was shown before the user \\ interacted with the system, e.g., recommending movies, products, or \\ books in online stores and streaming services before the user searches \\ for an item or interacts with the parameters.\end{tabular} \\ \midrule
9 &
  \begin{tabular}[c]{@{}l@{}}Others \\ (change)\end{tabular} &
  How have updates affected the system? &
  \begin{tabular}[c]{@{}l@{}}Asking how system updates have changed the functioning of the system, \\ e.g., new affordances, changed error boundaries, or updated feature \\ importances.\end{tabular} \\ \midrule
10 &
  \begin{tabular}[c]{@{}l@{}}Others\\ (change)\end{tabular} &
  Does the system learn specifically from my interaction? &
  \begin{tabular}[c]{@{}l@{}}Asking if the system is using this specific user's interaction to \\ change/adapt/improve the system over time, e.g., ``Is my conversation \\ with this chatbot being tracked and used for training?''.\end{tabular} \\ \midrule
11 &
  \begin{tabular}[c]{@{}l@{}}Others \\ (context)\end{tabular} &
  Who is responsible for this system? / Who are the authors? &
  \begin{tabular}[c]{@{}l@{}}Asking for background information on the system, \\ e.g., ``Who is the creator or author of the system?'' or ``Who is accountable?''.\end{tabular} \\ \bottomrule
\end{tabular}}
\caption{\label{tab:new-questions}Newly developed questions for identifying end-user explanation needs, along with a definition and example, as an extension to the XAIQB.}
\end{table*}

\section{Discussion}
%
In general, we find the XAIQB to be a useful tool for exploring user needs. 
%
Nevertheless, using the XAIQB for our deductive coding approach on our highly context-specific software exploration data, we noticed a lack of questions in the \emph{UI}, \emph{Output}, and \emph{Other} categories, as our study participants repeatedly expressed their need for explanation in these areas. 
Our findings echo Kim et al.'s~\cite{kim2023} empirically, which show that all users, including lay people, expressed a need for practically useful information to advance their human-AI collaboration, while it was mostly technical professionals who were interested in the details of the AI system. 
Therefore, instead of focusing solely on accurately explaining the technical inner workings of the AI system, we suggest that application- and user-specific explanation needs should be integrated into the XAIQB and addressed in the XAI system design. This is where we see the difference between HC-XAI and XAI.
Regarding the \emph{UI} needs, it should be noted that Liao et al.~\cite{Liao2020} consider Amershi et al.'s~\cite{Amershi_2019_GuidelinesAI} usability guidelines for human-AI interaction as relevant to explainability. However, as these guidelines don't provide guidance on how to act on them, the authors did not consider these usability needs in their XAIQB development.
In our opinion, how or if explainability needs could be satisfied should not affect whether or not they are recognized as such. 
Our results indicate that usability needs are an essential part of the needs of end-users, which is why we introduced UI- and context-specific questions. Future work could further explore the connection between XAI and usability needs in human-AI collaboration.
%

Regarding the applicability of the XAIQB, we made the following observations.
First, it took a lot of time during our discussion sessions to differentiate the existing questions, \eg ``How does the system make predictions?'' and ``What is the system’s overall logic?'' were often mapped to the same user statements.
Moreover, terms like ``what kind'', ``overall logic'' and ``top rules'', which are used in many of the XAIQB's questions, are not self-explanatory. Other more technical terms like ``parameters'', ``instance'', and ``data'' are not specific enough, \eg hyper- or system parameters.
We recommend discussing the XAIQB, including definitions, with the team to create a general mutual understanding in the context of the AI system to be developed.
This may be beneficial to identify the explanation needs of end-users. 

A limitation of our study is the generalizability to other domains, as we used only art historians as participants in our study and, \eg not lay people. Further empirical testing with other types of end-users, especially in realistic task settings, may lead to new questions and different definitions. Furthermore, we have not yet tested the comprehensibility of our questions and definitions with third parties. Therefore, we consider our definitions to be a work in progress.

\section{Conclusion and Future Work}
Our findings suggest that the XAIQB still lacks coverage of explanation needs in the areas of user interface and practical use of the AI system. During the analysis, we also found that some of the questions are difficult to differentiate and the terms used are ambiguous. With this work, we propose 11 new questions to cover the explanation needs of our subject matter expert participants that were not previously addressed in the XAIQB. For ease of use and clarification, we also provide definitions and examples for all previous XAIQB questions and ours. While the XAIQB is a useful starting point, for practical use we recommend additional customization, such as team discussions about use-case specific definitions of the questions, which can be based on our proposed definitions. We also recommend considering the UI and contextual questions we have developed. Human-centered questions such as these appear to be essential for identifying end-user XAI needs. To make the XAIQB more generalizable and holistic, future research should investigate additional use cases. It would also be interesting to evaluate the usefulness of the XAIQB in the actual development process of an AI system.
Finally, a comprehensive evaluation of the XAIQB requires an exhaustive systematic literature review of articles citing the XAIQB, especially empirical work.


\begin{acks}
We thank the study participants and reviewers for their valuable and insightful comments. This work is supported by
the Federal Ministry of Education and Research (grant 16DHBKI025:
ENKIS), the state of Berlin, and the European Union.
\end{acks}

\bibliographystyle{ACM-Reference-Format}
\bibliography{bibliography}


\end{document}